\documentclass{article}
\usepackage{amsfonts}
\usepackage{amssymb}
\usepackage{amsmath}
\usepackage{cite}
\allowdisplaybreaks
\numberwithin{equation}{section}
\DeclareMathOperator{\sgn}{\rm sgn}

\DeclareMathOperator*{\res}{\rm res}

\DeclareMathOperator{\RR}{\mathbb{R}}

\DeclareMathOperator{\CC}{\mathbb{C}}
\DeclareMathOperator{\Go}{\mathcal{G}}
\DeclareMathOperator{\Lo}{\mathcal{L}}
\DeclareMathOperator{\No}{\mathcal{N}}

\DeclareMathOperator{\bk}{\mathbf{k}}

\DeclareMathOperator{\vhi}{\varphi}
\DeclareMathOperator{\la}{\lambda}
\DeclareMathOperator{\ka}{\varkappa}
\def\t#1{\widetilde{#1}}

\def\h#1{\widehat{#1}}
\def\ol#1{\overline{#1}}

\begin{document}

\title{Cauchy--Jost function and hierarchy of integrable equations}
\author{M. Boiti$^\dag$, F. Pempinelli$^\dag$, A. K. Pogrebkov$^\ddag$\\
$^\dag$\ EINSTEIN Consortium, Lecce, Italy\\
$^\ddag$\ Steklov Mathematical Institute, Moscow, Russia}
\maketitle

\begin{abstract}
 Properties of the Cauchy--Jost (known also as Cauchy--Baker--Akhiezer) function of the KPII equation are described. By means of the $\bar\partial$-problem for this function it is shown that all equations of the KPII hierarchy are given in a compact and explicit form, including equations on the Cauchy--Jost function itself, time evolutions of the Jost solutions and evolutions of the potential of the heat equation.
\end{abstract}


\section{ Introduction}
Hirota bilinear identity is known to generate hierarchies of the integrable equations, see e.g.\ \cite{MJD} for the hierarchy of the Kadomtsev--Petviashvili equation. On the other side there exists a problem in prooving of this identity, as it is based on assumption that the scattering data vanish outside some finite region on the complex plane. This condition, being valid for some classes of solutions, say, soliton ones, is inappropriate when one considers solutions rapidly decaying at space infinity, as it is shown below. Here we proof that one of the relations derived in \cite{BK1} is free from this defect and can be used for both applications: it is valid for the class of rapidly decaying potentials, but  under some additional conditions it leads to the Hirota bilinear identity as well. Taking that in \cite{BK1} (see also \cite{BK2} for some other integrable hierarchies) this relation was derived from the Hirota identity we get the closed formulation of the hierarchy of integrable equations in terms of the Cauchy--Jost function $F(x,\la,\mu)$. This function being the primitive of the product of the Jost and dual Jost solutions is the main object of our construction. This function is very well known and naturally appears in the theory of (binary) B\"acklund transformations, see \cite{MS}. In \cite{GO} this  function was studied from an algebro-geometric point of view and it was called there the Cauchy--Baker--Akhiezer kernel. In \cite{BK1} it was called the  Cauchy--Baker--Akhiezer function. Here, since our consideration is based on the standard $\bar\partial$-approach for decaying solutions, we prefer to use the term  Cauchy--Jost function.

The article is organized as follows. In Sec.\ 2 we, following the standard formulation of the inverse problem for the heat equation in the case of decaying potential \cite{ip1,ip2}, we give necessary details of this approach and define the Cauchy--Jost function in terms of the Jost solutions. We also consider the problem of the proof of the Hirota bilinear identity specific for the case of decaying potentials. In Sec.\ 3 we present some properties of  the Cauchy--Jost function including $\bar\partial$-problem for it. In Sec.\ 4 we switch to an arbitrary number of ``times'' and consider the action of an operator generating these evolutions on the Cauchy--Jost function. This enables us to present in a compact form, see (\ref{as10}), all evolutions corresponding to the KP hierarchy. We get in this way relation (\ref{as10}) that coincides with Eq.\ (41) from \cite{BK1} with accuracy to that the latter one was derived there from the Hirota bilinear identity. In Sec.\ 5 we show that this relation generates not only hierarchy of evolutions, but  evolutions of the Jost solutions and corresponding second operators of the Lax pairs as well. Concluding remarks are given in the Sec.\ 6.

\section{Jost solutions, $\bar\partial$-problem for the heat equation and the Cauchy--Jost function}
We start with some standard properties of the Jost solutions of the heat equation, see, e.g., \cite{ip1,ip2} and \cite{BPPPr2001a} for details. We consider here Lax operator, and its dual
\begin{equation}
\mathcal{L}(x,\partial _{x}^{})=-\partial _{x_{2}}^{}+\partial _{x_{1}}^{2}+2w_{x_1}(x),\qquad \mathcal{L}^{\text{d}}(x,\partial _{x}^{})=\partial _{x_{2}}^{}+\partial _{x_{1}}^{2}+2w_{x_1}(x),\quad x=(x_1,x_2),
\label{old3}
\end{equation}
where for simplicity of the relations below we introduced function $w(x)$, related to the standard potential and $\tau$-function by means of
\begin{equation}
u(x)=-2w_{x_1}(x),\text{ so that }w(x)=\partial_{x_1}\ln\tau(x).\label{old4}
\end{equation}

Let $u(x)$ rapidly decays at space infinity, i.e., faster than $(x_{1}^{2}+x_{2}^{2})^{-1}$. Then the Jost solutions $\vhi(x,\la)$ and $\psi(x,\la)$
\begin{equation}
\mathcal{L}(x,\partial _{x}^{})\vhi(x,\la)=0,\qquad \mathcal{L}^{\text{d}}(x,\partial _{x}^{})\psi(x,\la)=0,\label{old41}
\end{equation}
of the heat equation and its dual are fixed by condition that functions $\chi(x,\la)$ and $\xi(x,\la)$
\begin{equation}
\vhi(x,\la)=e^{\ell(\la)x}\chi(x,\la),\qquad\psi(x,\la)=e^{-\ell(\la)x}\xi(x,\la), \label{old42}
\end{equation}
are bounded on the complex plane of the spectral parameter $\la$ for all $x\in\RR^{2}$ and
\begin{equation}
\lim_{\la\to\infty}\chi(x,\la)=\lim_{\la\to\infty}\xi(x,\la)=1,\qquad\lim_{x\to\infty}\chi(x,\la)=\lim_{x\to\infty}\xi(x,\la)=1.\label{old421}
\end{equation}
Here we denoted
\begin{equation}
\ell(\la)x=\la x_1+\la^{2} x_2.\label{old43}
\end{equation}

These functions obey
\begin{equation}
\bigl(\mathcal{L} (x,\partial_{x}^{})+2\la\partial_{x_{1}}\bigr)\chi (x,\la)=0,\qquad\bigl(\mathcal{L}^{\text{d}}(x,\partial_{x}^{})-2\la\partial_{x_{1}}\bigr)\xi(x,\la)=0,\label{old431}
\end{equation}
and are defined by the integral equations
\begin{align}
&\chi (x,\la) =1-2\!\!\int \!\! d x'\,G_{0}(x-x',\la)w_{x'_1}(x')\chi (x',\la),\label{old432} \\
&\xi (x,\la) =1-2\!\!\int \!\! dx'\,\xi (x',\la)w_{x'_1}(x')G_{0}(x'-x,\la),\label{old433}
\end{align}
where $G_{0}(x,\la)$ is the Green's function
\begin{align}
&\bigl(\mathcal{L}_{0}(x,\partial _{x}^{})+2\la\partial_{x_{1}}\bigr)G_{0}(x,\la)=\delta (x_{1})\delta (x_{2})\label{old434}\\
&G_{0}(x,\la)=\dfrac{1}{2\pi}\int d\alpha[\theta(\la^{2}_{\Im}-\alpha^{2})-\theta(x_{2})]e^{(\alpha-\la_{\Im})(ix_1-(\alpha+\la_{\Im}-2i\la_\Re)x_2)},\label{old435}
\end{align}
and $\Lo_{0}(x,\partial _{x}^{})$ is (\ref{old3}) with $w\equiv0$.  Relations (\ref{old432}) and (\ref{old433}) give asymptotic expansions for the Jost solutions:
\begin{align}
&\chi(x,\la)=\sum_{n=0}^{\infty}\biggl(\dfrac{\partial^{-1}_{x_1}\Lo(x,\partial _{x}^{})}{-2\la}\biggr)^{n}1
\equiv1-\dfrac{1}{\la}\sum_{n=0}^{\infty}\biggl(\dfrac{\partial^{-1}_{x_1}\Lo(x,\partial _{x}^{})}{-2\la}\biggr)^{n}w(x),\label{old436}\\
&\xi(x,\la)=\sum_{n=0}^{\infty}\biggl(\dfrac{\partial^{-1}_{x_1}\Lo^{\text{d}}(x,\partial _{x}^{})}{2\la}\biggr)^{n}1\equiv1+\dfrac{1}{\la}
\sum_{n=0}^{\infty}\biggl(\dfrac{\partial^{-1}_{x_1}\Lo^{\text{d}}(x,\partial _{x}^{})}{2\la}\biggr)^{n}w(x).\label{old437}
\end{align}
Here $\partial^{-1}_{x_1}$ denotes the inverse operator. Detailed study of asymptotics of (\ref{old435}) proves that this operator acts on a function $f(x)$ as
\begin{equation}
(\partial^{-1}_{x_1}f)(x)=\dfrac{\sgn\la_{\Im}}{2\pi{i}}\int d^{2}x'\dfrac{f(x')}{x_2-x'_2-i(x_1-x'_{1})\la_{\Im}0},\label{old438}
\end{equation}
generating under time evolution so called Manakov constraints (cf.\ \cite{KPI}, where this study was performed for the case of KPI equation). Here we skip these details.

In \cite{BPPPr2001a} we proved that the Jost solutions are given in terms of the total Green's function as
\begin{align}
&\vhi (x,\la) =\!\!\int \!\! d x'\,e^{\ell(\la)x'}\mathcal{L}_{0}^{\mathrm{d} }(x',\partial_{x'}^{})\Go(x,x',\la)\equiv
e^{\ell(\la)x}-2\!\!\int \!\! d x'\,e^{\ell(\la)x'}w_{x'_1}(x')\Go(x,x',\la),\label{old61} \\
&\psi (x',\la) =\!\!\int \!\! dx\,e^{-\ell(\la)x}\mathcal{L}_{0}^{} (x,\partial_{x}^{})\Go(x,x',\la)\equiv
e^{-\ell(\la)x'}-2\!\!\int \!\! dx\,e^{-\ell(\la)x}w_{x_1}(x)\Go(x,x',\la),\label{old71}\\
&\chi (x,\la) =\!\!\int \!\! d x'\,\bigl(\mathcal{L}_{0}^{\mathrm{d} }(x',\partial_{x'}^{})-2\la\partial_{x'_{1}}\bigr)G(x,x',\la)\equiv
1-2\!\!\int \!\! d x'\,G(x,x',\la)w_{x'_1}(x'),\label{old6} \\
&\xi (x',\la) =\!\!\int \!\! dx\,\bigl(\mathcal{L}_{0}^{} (x,\partial_{x}^{})+2\la\partial_{x_{1}}\bigr)G(x,x',\la)\equiv
1-2\!\!\int \!\! dx\,w_{x_1}(x)G(x,x',\la),\label{old7}
\end{align}
where
\begin{equation}
G(x,x',\la)=e^{\ell(\la)(x'-x)}\Go(x,x',\la),\label{old72}
\end{equation}
and where the Green's functions obey the differential equations
\begin{align}
&\mathcal{L}(x,\partial _{x}^{})\Go(x,x',\la)= \mathcal{L}_{}^{\mathrm{d}}(x',\partial_{x'}^{})\Go(x,x', \la)=\delta (x-x')\equiv\delta(x_1-x_1')\delta(x_2-x_2'),\label{old13}\\
&\bigl(\mathcal{L}(x,\partial _{x}^{})+2\la\partial_{x_{1}}\bigr)G(x,x',\la)=\bigl(\mathcal{L}_{}^{\mathrm{d}}(x',\partial_{x'}^{})-2\la\partial_{x'_{1}}\bigr)G(x,x', \la)=\delta (x-x').
\label{old131}
\end{align}
The second equalities in (\ref{old61})--(\ref{old7}) follow by (\ref{old3}),  (\ref{old13}) and (\ref{old131}). Relations (\ref{old6}) and (\ref{old7}) follow from (\ref{old61}), (\ref{old71}) and (\ref{old72}). Let us mention that  (\ref{old6}) and (\ref{old7}) give an exact meaning to decompositions  (\ref{old436}) and (\ref{old437}). The Green's functions in their turn are given by means of the Jost solutions:
\begin{align}
\Go(x,x',\la)&=\dfrac{1}{2\pi}\int d\alpha[\theta(\la^{2}_{\Im}-\alpha^{2})-\theta(x_{2}-x_{2}')]\vhi(x,\la_{\Re}+i\alpha)\psi(x',\la_{\Re}+i\alpha),\label{old8} \\
G(x,x',\la)&=\dfrac{1}{2\pi}\int d\alpha[\theta(\la^{2}_{\Im}-\alpha^{2})-\theta(x_{2}-x_{2}')]e^{(\alpha-\la_{\Im})(i(x_1-x'_1)-(\alpha+\la_{\Im}-2i\la_\Re)(x_2-x'_2))}\times\nonumber\\
&\times\chi(x,\la_{\Re}+i\alpha)\xi(x',\la_{\Re}+i\alpha).\label{old9}
\end{align}
Specific property of the Green's function follows from the observation that its functional derivative with respect to the potential is given in terms of the same function, see \cite{BPPPr2001a}:
\begin{equation}
\dfrac{\delta\Go(x,y,\la)}{\delta{u}(z)}=\Go(x,z,\la)\Go(z,y,\la),\label{old91}
\end{equation}
while functional derivatives of the Jost solutions involve the Green's function,
\begin{equation}
\dfrac{\delta\vhi(x,\la)}{\delta{u}(y)}=\Go(x,y,\la)\vhi(y,\la),\qquad\dfrac{\delta\psi(x,\la)}{\delta{u}(y)}=\psi(y,\la)\Go(y,x,\la),\label{old92}
\end{equation}
as follows by applying (\ref{old91}) to the first equalities in (\ref{old61}) and (\ref{old71}).

Thanks to the reality of the potential $w_{x_1}(x)$ we have the conjugation properties
\begin{equation}
\overline{\vhi (x,\la)}=\vhi (x,\bar{\la}),\qquad\overline{\psi (x,\la)}=\psi (x,\bar\la),\label{old14}
\end{equation}
and the normalization and completeness relations
\begin{align}
&\int \!\! d x_{1}\psi (x,\la+ip)\vhi (x,\la)=2\pi \delta (p),\quad p\in \mathbb{R},  \label{old15} \\
&\int\limits_{x_{2}'=x_{2}}\!\!\!\!\ d\la_{\Im }\psi (x', \la)\vhi(x,\la)=2\pi \delta (x_{1}-x_{1}'), \label{old16}
\end{align}
for the Jost solutions. In \cite{BPPPr2001a} we proved also that the total Green's function obeys
\begin{equation}
\dfrac{\partial{\Go(x,x',\la)}}{\partial\ol\la}=\dfrac{i\sgn\la_{\Im}}{2\pi}\vhi(x,\bar\la)\psi(x',\bar\la),\label{old17}
\end{equation}
that thanks to (\ref{old61}) and (\ref{old71}) provide us with the standard equations of the inverse problem:
\begin{align}
&\dfrac{\partial\vhi(x,\la)}{\partial\ol\la}=\vhi(x,\bar\la)r(\la),\qquad \dfrac{\partial\psi(x,\la)}{\partial\ol\la}=-\psi(x,\bar\la)r(\bar\la),\label{old19}\\
\intertext{or}
&\dfrac{\partial\chi(x,\mu)}{\partial\ol\mu}=\chi(x,\bar\mu)r(\mu)e^{(\ell(\bar\mu)-\ell(\mu))x},\qquad\dfrac{\partial\xi(x,\la)}{\partial\ol\la}=-r(\bar\la)e^{(\ell(\la)-\ell(\bar{\la}))x}\xi(x,\bar\la).\label{old18}
\end{align}
where $r(\la)$ is the scattering data
\begin{align}
&r(\la)=\dfrac{\sgn\la_{\Im}}{\pi{i}}\int d^{2}x\,e^{-\ell(\bar\la)x}w_{x_1}(x)\vhi(x,\la)\equiv\dfrac{\sgn\la_{\Im}}{\pi{i}}\int d^{2}x\,e^{2i\Im\ell(\la)x}w_{x_1}(x)\chi(x,\la),\label{old21}\\
&r(\la)=\dfrac{\sgn\la_{\Im}}{\pi{i}}\int d^{2}x\,e^{\ell(\la)x}w_{x_1}(x)\psi(x,\bar\la)\equiv\dfrac{\sgn\la_{\Im}}{\pi{i}}\int d^{2}x\,e^{2i\Im\ell(\la)x}w_{x_1}(x)\xi(x,\bar\la),\label{old22}
\end{align}
that obeys conjugation property
\begin{equation}
\ol{r(\la)}=r(\bar\la).\label{old20}
\end{equation}
Here we assume that equations (\ref{old18}) are uniquely solvable under condition
\begin{equation}
\lim_{\la\to\infty}\chi(x,\la)=1,\text{ and }\lim_{\la\to\infty}\xi(x,\la)=1,\label{old23}
\end{equation}
correspondingly.

The Hirota bilinear identity, see e.g.\ \cite{MJD}, in these terms has the form
\begin{equation}
\oint_{C} d\la\vhi(x,\la)\psi(x',\la)=0,\label{old24}
\end{equation}
where $C$ is an arbitrary (smooth) contour in vicinity of infinity on the complex $\la$-plane. The standard derivation of this equality is based on assumption on dependence of $\vhi(x,\la)$ and $\psi(x',\la)$ on infinite sets of variables $x=(x_1,x_2,\ldots)$, $x'=(x'_1,x'_2,\ldots)$, in particular by substitution of a formal series
\begin{equation}
\ell(\la)x=\sum_{k=1}^{\infty}\la^{k}x_k,\label{as}
\end{equation}
for (\ref{old43}) in (\ref{old42}). In \cite{MJD} it is shown that  (assuming applicability of asymptotic decompositions (\ref{old436}) and (\ref{old437}) under the sign of the integral) one derives,  by means of formal algebraic operations, that potential $u(x)\equiv{u}(x_1,x_2,x_3,x_4,\ldots)$ obeys equations of the KPII hierarchy with respect to all $x_k$. Not going into details of definition of the functions of infinite set of variables, let us show that derivation of  the Hirota identity (\ref{old24}) in the case of decaying potential $u(x)$ has a specific problem. Indeed, by (\ref{old19}) and the Green's formula we have that
\begin{align*}
\oint_{C}d\la\,\vhi(x,\la)\psi(x',\la)&=\int_{D}d\bar\la\wedge{d}\la\,\dfrac{\partial}{\partial\bar\la}\vhi(x,\la)\psi(x',\la)=\\
&=\biggl(\int_{D}-\int_{\bar{D}}\biggr)d\la\wedge{d}\bar{\la}\,r(\la)\vhi(x,\bar\la)\psi(x',\la),
\end{align*}
where $D$ is interior of the contour $C$ on the complex $\la$-plane an $\bar{D}=\{\la\in\CC|\bar\la\in{D}\}$. Thus the r.h.s.\ equals to zero only if domain $D$ (and then the contour $C$) is symmetric with respect to the real $\la$-axis. The only possibility to get zero for an arbitrary contour at infinity is to impose the additional condition that $r(\la)$ has a finite support belonging to $D\cap\bar{D}$, see \cite{BK1} for instance. On the other side the generating function of integrals of motion in the case of a decaying potential is known (see, e.g.\  \cite{BPPPr2001a}) to be equal to
\begin{equation}
\rho(\la)=\int d^{2}x\,w_{x_1}(x)\chi(x,\la)\equiv\int d^{2}x\,w_{x_1}(x)\xi(x,\bar\la),\label{old25}
\end{equation}
so that on the real axis, because of (\ref{old21}) and (\ref{old22}), we have
\begin{equation}
\lim_{\la_{\Im}\to\pm0}r(\la)=\dfrac{\pm1}{\pi{i}}\rho(\la_{\Re}).\label{old26}
\end{equation}
Thus the assumption that $r(\la)$ has a finite support implies that all integrals of motion (coefficients of $1/\la$ expansion of $\rho(\la)$) equal zero. In other words, it results in trivial solutions of KPII. Let us underline, that condition of decaying of potential is essential here. Say, in the case of soliton potential the Jost solutions obey (\ref{old24}), since the scattering data for soliton potentials are finite sums of $\delta$-functions, i.e., they are finitely supported. At the same time the argumentation as above is not applicable here because for the soliton solutions the integrals of motion (\ref{old25}) are divergent in this case, \cite{KPII}.

Here we demonstrate that the consideration based on the Cauchy--Jost function, defined in terms of the Jost solutions as
\begin{equation}
F(x,\la,\mu)=\!\!\!\!\!\!\int\limits^{x_1}_{\substack{(\la-\mu)_{\Re}\infty \\y_2=x_2}}\!\!\!\!\!\! dy_{1}\psi(y,\la)\vhi(y,\mu),\label{g15}
\end{equation}
where $(\la-\mu)_{\Re}\infty$ denotes the sign of the infinite limit, gives an explicit description of the whole KPII hierarchy, including equations themselves and time evolutions of the Jost solutions. In the next section we consider properties of this function in detail.

\section{The Cauchy--Jost function in the case of a rapidly decaying potential}
The integral in (\ref{g15}) is convergent for all $\la\neq\mu$ as follows from asymptotic behavior (\ref{old42})--(\ref{old43}) of the Jost solutions and defines for these values of $\la,\mu\in\CC$ a smooth function of its arguments. At $\mu=\la$ the Cauchy--Jost function has a pole behavior \cite{GO}, because by (\ref{g15})
\begin{equation}
\lim_{\la_{\Re}\to\mu_{\Re}\pm0}F(x,\la,\mu)=\!\!\!\!\!\!\int\limits^{x_1}_{\substack{\pm\infty \\y_2=x_2}}\!\!\!\!\!\! dy_{1}\psi(y,\mu_{\Re}+i\la_{\Im})\vhi(y,\mu),\label{g16}
\end{equation}
and orthogonality relation (\ref{old15}) for the Jost solutions gives
\begin{equation}
\bigl(\lim_{\la_{\Re}\to\mu_{\Re}+0}-\lim_{\la_{\Re}\to\mu_{\Re}-0}\bigr)F(x,\la,\mu)=-2\pi\delta(\la_{\Im}-\mu_{\Im}).\label{g17}
\end{equation}
Thus $\res_{\mu=\la}F(x,\la,\mu)=1$.

In order to control the asymptotic behavior of this function we introduce, in analogy to (\ref{old42}), the Cauchy--Jost function with removed exponential factors:
\begin{equation}
f(x,\la,\mu)=e^{(\ell(\la)-\ell(\mu))x}F(x,\la,\mu).\label{g232}
\end{equation}
Then thanks to (\ref{old42}) equality (\ref{g15}) gives
\begin{equation}
f(x,\la,\mu)=\!\!\!\!\!\!\int\limits^{x_1}_{\substack{(\la-\mu)_{\Re}\infty \\y_2=x_2}}\!\!\!\!\!\! dy_{1}e^{(\la-\mu)(x_1-y_1)}\xi(y,\la)\chi(y,\mu),\label{g233}
\end{equation}
that under the asymptotic condition (\ref{old421}) on the Jost solutions shows that
\begin{equation}
f(x,\la,\mu)\to0\text{ when either }\la\to\infty,\text{ or }\mu\to \infty.\label{g234}
\end{equation}
Let us also introduce the function $g(x,\la,\mu)$ by means of the equality
\begin{equation}
f(x,\la,\mu)=\dfrac{1}{\mu-\la}+g(x,\la,\mu).\label{g8}
\end{equation}
Thanks to (\ref{g233}) it equals
\begin{equation}
g(x,\la,\mu)=\!\!\!\!\!\!\int\limits^{x_1}_{\substack{(\la-\mu)_{\Re}\infty \\y_2=x_2}}\!\!\!\!\!\! dy_{1}e^{(\la-\mu)(x_1-y_1)}\bigl(\xi(y,\la)\chi(y,\mu)-1\bigr),\label{g2331}
\end{equation}
and by (\ref{old42}) this function is bounded for all $x\in\RR^2$ and any $\la,\mu\in\mathbb{C}$ such that $\la\neq\bar\mu$. This function decays when $x\to\infty$, or when either $\la$ or $\mu$ tends to infinity. Again thanks to (\ref{old42}) and (\ref{old421}) we have for the asymptotic behavior of $f(x,\la,\mu)$:
\begin{align}
\lim_{\la\to\infty}\la f(x,\la,\mu)&=-1+\lim_{\la\to\infty}\la g(x,\la,\mu)=-\chi(x,\mu),\label{g5}\\
\lim_{\mu\to\infty}\mu f(x,\la,\mu)&=1+\lim_{\mu\to\infty}\mu g(x,\la,\mu)=\xi(x,\la),\label{g6}
\end{align}
and thanks to the leading terms of the asymptotic expansions (\ref{old436}), (\ref{old437}):
\begin{equation}
\lim_{\substack{\la\to\infty\\ \mu\to\infty}}\la\mu g(x,\la,\mu)=w(x),\label{g7}
\end{equation}
or by (\ref{old4})
\begin{equation}
u(x)=-\dfrac{1}{2}\lim_{\substack{\la\to\infty\\ \mu\to\infty}}\la\mu f_{x_1}(x,\la,\mu).\label{g231}
\end{equation}

Taking singular behavior of $F(x,\la,\mu)$ at $\la=\mu$ and relations (\ref{g15}) and (\ref{old19}) into account we get the $\bar\partial$-derivatives of the Cauchy--Jost function with respect to the spectral parameters:
\begin{equation}
\dfrac{\partial F(x,\la,\mu)}{\partial\ol\la}=-\pi\delta(\la-\mu)-r(\bar\la)F(x,\bar\la,\mu),\qquad \dfrac{\partial F(x,\la,\mu)}{\partial\ol\mu}=\pi\delta(\la-\mu)+F(x,\la,\bar\mu)r(\mu),\label{g20}
\end{equation}
where here and below the $\delta$-function of a complex variable is defined as
\begin{equation}
\delta(\la)=\delta(\la_{\Re})\delta(\la_{\Im}),\label{g201}
\end{equation}
so that $\partial_{\bar\la}1/\la=\pi\delta(\la)$, $\la\in\CC$, and ${\int}d\bar{z}\wedge{dz}\,\delta(z)=2i$. In terms of function $f(x,\la,\mu)$ Eq.\ (\ref{g20}) sounds as
\begin{align}
&\dfrac{\partial f(x,\la,\mu)}{\partial\bar\la}=-\pi\delta(\la-\mu)-r(\bar\la)e^{(\ell(\la)-\ell(\bar{\la}))x}f(x,\bar\la,\mu),\label{g24}\\
&\dfrac{\partial f(x,\la,\mu)}{\partial\bar\mu}=\pi\delta(\la-\mu)+f(x,\la,\bar\mu)r(\mu)e^{(\ell(\bar\mu)-\ell(\mu))x}.\label{g25}
\end{align}
Taking asymptotic behavior of $f(x,\la,\mu)$ into account these equations can be reduced to any of the following integral equations:
\begin{align}
f(x,\la,\mu)&=\dfrac{1}{\mu-\la}+\dfrac{1}{2\pi{i}}\int\dfrac{d\bar{z}{\wedge}dz\,r(\bar{z})}{z-\la}e^{(\ell(z)-\ell(\bar{z}))x}f(x,\bar{z},\mu),\label{g26}\\
f(x,\la,\mu)&=\dfrac{1}{\mu-\la}-\dfrac{1}{2\pi{i}}\int\dfrac{d\bar{z}{\wedge}dz\,r(z)}{z-\mu}e^{(\ell(\bar{z})-\ell(z))x}f(x,\la,\bar{z}),\label{g27}\\
f(x,\la,\mu)&=\dfrac{1}{\mu-\la}-\dfrac{1}{2\pi{i}}\int\dfrac{d\bar{z}{\wedge}dz\,r(z)}{(\la-\bar{z})(\mu-z)}e^{(\ell(\bar{z})-\ell(z))x}-\nonumber\\
&-\dfrac{1}{(2\pi{i})^2}\int\dfrac{d\bar{z}{\wedge}dz\,r(\bar{z})}{\la-z}\int\dfrac{d\bar{z'}{\wedge}dz'\,r(z')}{\mu-z'}e^{(\ell(z)-\ell(\bar{z})+\ell(\bar{z'})-\ell(z'))x}
f(x,\bar{z},\bar{z'}).\label{g28}
\end{align}
In particular, functions $f(x,\la,\mu)$ and  $F(x,\la,\mu)$ corresponding to the zero scattering data equal
\begin{equation}
f_{0}(x,\la,\mu)=\dfrac{1}{\mu-\la},\qquad F_{0}(x,\la,\mu)=\dfrac{e^{(\ell(\mu)-\ell(\la))x}}{\mu-\la},\label{g281}
\end{equation}
where the second equality follows from (\ref{g232}).

Thanks to the orthogonality relation (\ref{old15}) we derive from (\ref{g16}) that
\begin{equation}
\lim_{x_{1}\to\pm\infty}\lim_{\la_{\Re}\to\mu_{\Re}\pm0}F(x,\la,\mu)=0,
\qquad\lim_{x_{1}\to\mp\infty}\lim_{\la_{\Re}\to\mu_{\Re}\pm0}F(x,\la,\mu)=\mp2\pi\delta(\la_{\Im}-\mu_{\Im}).\label{g171}
\end{equation}
Then by (\ref{g20})
\begin{align}
&\lim_{x_{1}\to\mp\infty}\lim_{\la_{\Re}\to\mu_{\Re}\pm0}\dfrac{\partial F(x,\la,\mu)}{\partial\bar\la}=\pm2\pi{r}(\bar\la)\delta(\la_{\Im}+\mu_{\Im}),\label{g21}\\
&\lim_{x_{1}\to\mp\infty}\lim_{\la_{\Re}\to\mu_{\Re}\pm0}\dfrac{\partial F(x,\la,\mu)}{\partial\bar\mu}=\mp2\pi{r}(\bar\la)\delta(\la_{\Im}+\mu_{\Im}).\label{g22}
\end{align}
Thus function  $F(x,\la,\mu)$ being given defines both: the primitive $w(x)$ of the potential $u(x)$ by means of the limiting procedure in (\ref{g7}), and scattering data by the previous equalities.

It is worth to mention that the whole construction can be inverted in the sense that one can start with the Cauchy--Jost function defined as decaying with respect to $\la$ (or $\mu$) solution of Eq.\ (\ref{g24}) (or (\ref{g25}) correspondingly), assuming that under this limiting condition Eq.\ (\ref{g24}) (or (\ref{g25})) is uniquely solvable under some smoothness and small norm conditions on the scattering data $r(\la)$. In order to prove that the solution of (\ref{g24}) obeys (\ref{g25}) as well, we consider the $\bar{\partial}_{\la}$ derivative of the difference of the left and right hand sides of (\ref{g25}):
\begin{align*}
&\dfrac{\partial}{\partial\bar\la}\Biggl(\dfrac{\partial f(x,\la,\mu)}{\partial\bar\mu}-\pi\delta(\la-\mu)-f(x,\la,\bar\mu)r(\mu)e^{(\ell(\bar\mu)-\ell(\mu))x}\Biggr)=\\
&\qquad\qquad=-r(\bar\la)e^{(\ell(\la)-\ell(\bar{\la}))x}\Biggl(\dfrac{\partial f(x,\bar\la,\mu)}{\partial\bar\mu}-\pi\delta(\bar\la-\mu)-f(x,\bar\la,\bar\mu)r(\mu)e^{(\ell(\bar\mu)-\ell(\mu))x}\Biggr),
\end{align*}
where (\ref{g24}) was used. We see that expression in parenthesis obeys homogeneous equation (\ref{g24}) and decays when $\la\to\infty$ because of condition on $f(x,\la,\mu)$. Because of the assumption of unique solvability of (\ref{g24}) the homogeneous equation has the zero solution only. This proves that Eq.\ (\ref{g25}) follows from (\ref{g24}). Opposite statement can be proved in the same way.  Let now the function $g(x,\la,\mu)$ be defined by (\ref{g8}). As consequence of (\ref{g24})  it obeys
\begin{equation}
\dfrac{\partial g(x,\la,\mu)}{\partial\bar\la}=-\dfrac{r(\bar\la)e^{(\ell(\la)-\ell(\bar{\la}))x}}{\mu-\bar\la}-r(\bar\la)e^{(\ell(\la)-\ell(\bar{\la}))x}g(x,\bar\la,\mu),\label{g29}
\end{equation}
or the integral equation
\begin{equation}
g(x,\la,\mu)=\dfrac{1}{2\pi{i}}\Biggl(-\int\dfrac{d\bar\nu{\wedge}d\nu\,r(\bar\nu)}{(\nu-\la)(\bar\nu-\mu)}e^{(\ell(\nu)-\ell(\bar\nu))x}+\int\dfrac{d\bar\nu{\wedge}d\nu\,r(\bar\nu)}{\nu-\la}e^{(\ell(\nu)-\ell(\bar\nu))x}g(x,\bar\nu,\mu)\Biggr),\label{g30}
\end{equation}
as consequence of (\ref{g26}). The inhomogeneous term here is a smooth bounded function of $x\in\RR^2$ and any $\la,\mu\in\CC$ with the exception of $\la=\bar\mu$, where it can have a logarithmic singularity $\sim\log|\la-\bar\mu|$. Below we assume that this is  the only singularity of $g(x,\la\mu)$.

Finally, if $f(x,\la,\mu)$ is given, one defines the Jost solutions and the potential of the heat equation by means of (\ref{g5}), (\ref{g6}) and (\ref{g231}), or (\ref{g7}). Let us also mention that because of (\ref{g30}) the function $g(x,\la,\mu)$ decays when $\la\to\infty$, or $\mu\to\infty$, admits asymptotic decompositions with respect to  both these complex variables and coefficients of the expansions do not depend on the order of these operations. Moreover, let us have two scattering data, $r(\la)$ and $r'(\la)$ and corresponding functions $F(x,\la,\mu)$ and $F'(x,\la,\mu)$ given by (\ref{g24}) or (\ref{g25}). Then thanks, say, to the first equation in (\ref{g20}) we get
\begin{align*}
\dfrac{\partial}{\partial\bar\la} [F'(x,\la,\mu)- F(x,\la,\mu)]=[r(\bar\la)-r'(\bar\la)]F'(x,\bar\la,\mu)-r(\bar\la)[F'(x,\bar\la,\mu)-F(x,\bar\la,\mu)],
\end{align*}
so that the difference $F'-F$ obeys the same integral equation with inhomogeneous term equal to $(2i)^{-1}\int{d\bar\nu}\wedge{d\nu}\delta(\la-\nu)[r(\bar\nu)-r'(\bar\nu)]F'(x,\bar\nu,\mu)$. This difference decays when $\la\to\infty$, so thanks to the assumption of unique solvability of (\ref{g20}) we derive that
\begin{equation}
F'(x,\la,\mu)-F(x,\la,\mu)=\dfrac{1}{2\pi{i}}\int {d\bar\nu}\wedge{d\nu}\,F'(x,\la,\bar\nu)(r'(\nu)-r(\nu))F(x,\nu,\mu).\label{as321}
\end{equation}
Taking (\ref{g281}) into account we see that this equality generalizes the integral equations on $F$ and reduces to them when either $r'(\la)$ or $r(\la)$ is identically zero. Also by means of this equality we have that
\begin{equation}
\dfrac{\delta F(x,\la,\mu)}{\delta r(\nu)}=F(x,\la,\bar{\nu})F(x,\nu,\mu),\label{as322}
\end{equation}
i.e., we have for the functional derivative of the Cauchy--Jost function with respect to the scattering data an equality similar to the functional derivative of the Green's function with respect to the potential, see (\ref{old91}). To calculate the functional derivative of $F(x,\la,\mu)$ with respect to $u(x)$ we write by (\ref{old92})
\begin{equation*}
\dfrac{\delta\psi(x,\la)\vhi(x,\mu)}{\delta{u}(y)}=\psi(y,\la)\Go(y,x,\la)\vhi(x,\mu)+\psi(x,\la)\Go(x,y,\mu)\vhi(y,\mu),
\end{equation*}
so that by (\ref{old8}) and (\ref{g15}) we derive
\begin{align}
\dfrac{\delta{F}(x,\la,\mu)}{\delta{u}(y)}&=\dfrac{1}{2\pi}\int d\alpha[\theta(\la^{2}_{\Im}-\alpha^{2})-\theta(y_{2}-x_{2})]F_{y_1}(y,\la,\la_{\Re}+i\alpha)F(x,\la_{\Re}+i\alpha,\mu)+\nonumber\\
&+\dfrac{1}{2\pi}\int d\alpha[\theta(\mu^{2}_{\Im}-\alpha^{2})-\theta(x_{2}-y_{2})]F(x,\la,\mu_{\Re}+i\alpha)F_{y_{1}}(y,\mu_{\Re}+i\alpha,\mu).\label{g31}
\end{align}
Relations (\ref{as322}) and (\ref{g31}) demonstrate a specific property of the Cauchy--Jost function: its infinitesimal transformations are given in terms of bilinear combinations of this function itself. Below we get a confirmation of this property by consideration of the time evolutions.

\section{Evolution with respect to an arbitrary number of times}

 Here we consider evolution of the potential of the heat operator (\ref{old3}) with respect to an infinite set of times $x=(x_1,x_2,x_3,\ldots)$. Strictly speaking, the role of times is played by variables $x_k$ with $k\geq3$, while variables $x_1$ and $x_2$ involved in the heat equation (\ref{old41}) are the space ones. In oder to avoid problem of definition of a functions depending on infinite number of variables we introduce here the dependence on an arbitrary finite number $N$ of times $x_1,\ldots,x_{N}$ setting
\begin{equation}
\ell(\la)x=\sum_{k=1}^{N}\la^{k}x_k,\label{lN}
\end{equation}
in all equations of the $\bar\partial$-problem (cf.\ (\ref{as})). Precisely, we substitute in (\ref{old43})  (thus, in fact, in (\ref{g15}) and in (\ref{g233}), where condition $x_2=y_2$ must be changed for $x_k=y_k$ for all $k\geq2$) $\ell(\la)x$ defined in (\ref{old42}) for   $\ell(\la)x$ as defined in (\ref{lN}). The same substitution must be performed in (\ref{g281}) and in equations (\ref{old18}), (\ref{g24}), (\ref{g25}), (\ref{g26})--(\ref{g28}), (\ref{g29}) and (\ref{g30}) that give the $\bar\partial$-derivatives of the Jost solutions and the Cauchy--Jost function. Then by (\ref{g24}) and (\ref{g25}) we have that
\begin{align}
&\dfrac{\partial}{\partial\ol\la}f_{x_k}(x,\la,\mu)=({\bar\la}^{k}-\la^{k})r(\bar\la)e^{(\ell(\la)-\ell(\bar\la))x}f(x,\bar\la,\mu)-r(\bar\la)e^{(\ell(\la)-\ell(\bar{\la}))x}f_{x_k}(x,\bar\la,\mu),\label{e1}\\
&\dfrac{\partial}{\partial\ol\mu}f_{x_k}(x,\la,\mu)=({\bar\mu}^{k}-\mu^{k})f(x,\la,\bar\mu)r(\mu)e^{(\ell(\bar\mu)-\ell(\mu))x}+f_{x_k}(x,\la,\bar\mu)r(\mu)e^{(\ell(\bar\mu)-\ell(\mu))x}.\label{e2}
\end{align}
for any $k\leq{N}$, while for $k>N$ the inhomogeneous terms in the r.h.s.'s equal to zero. Because of the assumption of unique solvability of these equations $f_{x_k}(x,\la,\mu)=0$ for all $k>N$. Thanks to (\ref{e1}) we have:
\begin{align}
\dfrac{\partial}{\partial\bar\la}\sum_{k=1}^{\infty}\dfrac{\partial_{x_k}}{\nu^{k+1}}f(x,\la,\mu)&=-\biggl(\dfrac{1-(\la/\nu)^{N+1}}{\nu-\la}-\dfrac{1-(\bar\la/\nu)^{N+1}}{\nu-\bar\la}\biggr)r(\bar\la)e^{(\ell(\la)-\ell(\bar{\la}))x}f(x,\bar\la,\mu)-\nonumber\\
&-r(\bar\la)e^{(\ell(\la)-\ell(\bar{\la}))x}\sum_{k=1}^{\infty}\dfrac{\partial_{x_k}}{\nu^{k+1}}f(x,\bar\la,\mu),\label{e3}
\end{align}
where sums are finite due to the remark above. Then the $k$-th equation in (\ref{e1}) is nothing but the equality of coefficients of $\nu^{-k-1}$ of the expansion with respect to $1/\nu$ of the left and right sides of this equality.

Let us define now an operator $X(\nu)$ by its action on the Cauchy--Jost function by means of the equality
\begin{align}
\dfrac{\partial}{\partial\bar\la}X(\nu)f(x,\la,\mu)&=-\biggl(\dfrac{1}{\nu-\la}-\dfrac{1}{\nu-\bar\la}\biggr)r(\bar\la)e^{(\ell(\la)-\ell(\bar{\la}))x}f(x,\bar\la,\mu)-\nonumber\\
&-r(\bar\la)e^{(\ell(\la)-\ell(\bar{\la}))x}X(\nu)f(x,\bar\la,\mu).\label{e4}
\end{align}
It is clear that the $1/\nu$-expansion of the inhomogeneous term coincides with the expansion of the inhomogeneous term in (\ref{e3}) up to $\nu^{-N-1}$. Thus  $X(\nu)f(x,\la,\mu)$ admits $1/\nu$ decomposition as well, and coefficient of $\nu^{-k-1}$ equals $f_{x_k}(x,\la,\mu)$ for all $k\leq{N}$. Taking into account  that (\ref{lN}) is a special case of (\ref{as}), where $x_k=0$ for all $k>N$,  we get that for any for $k>N$ this coefficient equals $f_{x_k}(x,\la,\mu)|_{x_{j}=0,\,j>N}$. This formulation enables to avoid the formal approach based on the usage of functions of infinite number of variables, where $\ell(\la)x$ is introduced by means of (\ref{as}) and operator $X(\nu)$  by equality
\begin{align}
X(\nu)=\sum_{k=0}^{\infty}\dfrac{\partial_{x_k}}{\nu^{k+1}},\quad\nu\in\CC,\quad\nu\neq0.\label{as0}
\end{align}
Because of the arbitrariness of $N$ in (\ref{lN}), the operator  $X(\nu)$ introduced in (\ref{e4}) can be considered a generating operator for the derivatives of the Cauchy--Jost function with respect to all higher times.

Using (\ref{g24}) we rewrite now equation (\ref{e4}) as
\begin{align*}
\dfrac{\partial}{\partial\bar\la}\biggl(X(\nu)f(x,\la,\mu)-\dfrac{f(x,\la,\mu)}{\nu-\la}\biggr)&=\pi\dfrac{\delta(\la-\mu)}{\nu-\mu}+\pi\delta(\la-\nu)f(x,\nu,\mu)-\\
&-r(\bar\la)e^{(\ell(\la)-\ell(\bar{\la}))x}\biggl(X(\nu)f(x,\bar\la,\mu)-\dfrac{f(x,\bar\la,\mu)}{\nu-\bar\la}\biggr),
\end{align*}
so that the expression in parenthesis obeys a  $\bar\partial$-equation  (\ref{g24}) with the inhomogeneous term
\begin{equation*}
\dfrac{\delta(\la-\mu)}{\nu-\mu}+\delta(\la-\nu)f(x,\nu,\mu)=\int \dfrac{d\bar\nu'\wedge{d}\nu'}{2i}\delta(\la-\nu')\biggl(\dfrac{\delta(\nu'-\mu)}{\nu-\mu}+\delta(\nu'-\nu)f(x,\nu,\mu)\biggr).
\end{equation*}
In the sense of distributions this term decays when $\la\to\infty$, so that because of (\ref{g24}) we get
\begin{equation*}
X(\nu)f(x,\la,\mu)-\dfrac{f(x,\la,\mu)}{\nu-\la}=-\int \dfrac{d\bar\nu'\wedge{d}\nu'}{2i}f(x,\la,\nu')\biggl(\dfrac{\delta(\nu'-\mu)}{\nu-\mu}+\delta(\nu'-\nu)f(x,\nu,\mu)\biggr),
\end{equation*}
that after integration in the r.h.s.\ gives
\begin{equation}
X(\nu)f(x,\la,\mu)=\biggl(\dfrac{1}{\mu-\nu}-\dfrac{1}{\la-\nu}\biggr)f(x,\la,\mu)-f(x,\la,\nu)f(x,\nu,\mu).\label{as9}
\end{equation}

Now thanks to  (\ref{g8}) and asymptotic properties (\ref{g5})--(\ref{g7}) we get action of operator $X(\nu)$ on the Jost solutions and potential:
\begin{align}
&X(\nu)\chi(x,\mu)=\dfrac{\chi(x,\mu)}{\mu-\nu}-\chi(x,\nu)f(x,\nu,\mu),\label{h24}\\
&X(\nu)\xi(x,\la)=-\dfrac{\xi(x,\la)}{\la-\nu}-f(x,\la,\nu)\xi(x,\nu),\label{h25}\\
&X(\nu)w(x)=\chi(x,\nu)\xi(x,\nu)-1\equiv\vhi(x,\nu)\psi(x,\nu)-1,\label{h28}
\end{align}
where (\ref{old42}) was used in the last equality. Evolution of the function $f(x,\la,\mu)$, Jost solutions and potential $w(x)$ with respect to any finite number of times of the KPII hierarchy follows by means of $1/\nu$ expansion of relations (\ref{as9})--(\ref{h28}).

Eq.\ (\ref{as9}) deserves some comments. Its l.h.s.\ is regular for any values $\la,\mu,\nu$ because the only singularity at $\la=\mu$ cancels out under differentiation. In order to check the regularity of the r.h.s.\ one can use (\ref{g8}) and rewrite (\ref{as9}) as
\begin{equation}
X(\nu)f(x,\la,\mu)=\dfrac{g(x,\la,\mu)-g(x,\la,\nu)}{\mu-\nu}-\dfrac{g(x,\la,\mu)-g(x,\nu,\mu)}{\la-\nu}-g(x,\la,\nu)g(x,\nu,\mu),\label{h29}
\end{equation}
so that the singularities at $\nu=\la$ and $\nu=\mu$ also cancel out, while the limiting values at these points can depend on the way of the limiting procedure. It is easy to see that operators $X(\nu)$ and $X(\nu')$ commute. Indeed, thanks to, say (\ref{as0}), we have
\begin{align}
X(\nu')&X(\nu)f(x,\la,\mu)=\biggl(\dfrac{1}{\mu-\nu'}-\dfrac{1}{\la-\nu'}\biggr)\biggl(\dfrac{1}{\mu-\nu}-\dfrac{1}{\la-\nu}\biggr)f(x,\la,\mu)-\nonumber\\
&-\biggl(\dfrac{1}{\mu-\nu}-\dfrac{1}{\la-\nu}\biggr)f(x,\la,\nu')f(x,\nu',\mu)-\biggl(\dfrac{1}{\mu-\nu'}-\dfrac{1}{\la-\nu'}\biggr)f(x,\la,\nu)f(x,\nu,\mu)+\nonumber\\
&+f(x,\la,\nu')f(x,\nu',\nu)f(x,\nu,\mu)+f(x,\la,\nu)f(x,\nu,\nu')f(x,\nu',\mu).\label{h30}
\end{align}
The same is valid, of course, for (\ref{h24})--(\ref{h28}).

Let us notice that, substituting in (\ref{old42}) $\ell(\la)x$ for a formal series in (\ref{as}), assuming that $|\nu|>|\la|,|\mu|$ and using the operator $X(\nu)$ given in (\ref{as0}), one can formally write equations  (\ref{as9})--(\ref{h25}) in a more compact form
\begin{align}
&X(\nu)F(x,\la,\mu)=-F(x,\la,\nu)F(x,\nu,\mu).\label{as10}\\
&X(\nu)\vhi(x,\mu)=-\vhi(x,\nu)F(x,\nu,\mu),\label{h26}\\
&X(\nu)\psi(x,\la)=-F(x,\la,\nu)\psi(x,\nu).\label{h27}
\end{align}
Equality  \ref{as10} demonstrates once more (cf.\ (\ref{as322}) and (\ref{g31})) that the infinitesimal transformations of the Cauchy--Jost function are given in terms of its bilinear combinations. Equations   (\ref{as10})--(\ref{h27}) were derived in \cite{BK1} in the case of a finitely supported potential. It was shown there that relation (\ref{as10}) enables the derivation of all equations of the KPII and, mKP hierarchies. In the next section we consider some specific aspects of the asymptotic expansion of this equality. Here let us mention that under the same assumption on finite support of the scattering data as in the derivation of the Hirota identity (\ref{old24}), we get by (\ref{g20})
\begin{equation}
\oint_{C} d\nu\, F(x,\la,\nu)F(x',\nu,\mu)=0,\label{h31}
\end{equation}
for $\la$ and $\mu$ outside the contour $C$. This equality in its turn leads to the Hirota identity because of the asymptotic properties of the Cauchy--Jost function. As we proved here, the assumption on the support of the scattering data $r(\la)$ is not necessary in order to get relations (\ref{as9})--(\ref{h28}). At the same time, relations (\ref{h24}) and (\ref{h25}) (or their formal counterparts (\ref{h26}) and (\ref{h27})) demonstrate that the Cauchy--Jost function naturally appears in the study of the evolution of the Jost solutions.

\section{Asymptotic series}
As follows from the above construction, the derivative of the Cauchy--Jost function $f(x,\la,\mu)$ with respect to some time $x_k$ appears as a coefficient of the term $\nu^{-k-1}$ of the asymptotic expansion of (\ref{as9}) with respect to $1/\nu$, $\nu\in\CC$. In the r.h.s.\ of this equality parameter $\nu$ appears explicitly and as argument of functions  $f(x,\la,\nu)$ and $f(x,\nu,\mu)$, so we need asymptotic expansions of the function $f(x,\la,\mu)$ with respect to both, $\la$ and $\mu$. But as follow from (\ref{g8}) these expansions depend on their order because of the term $(\mu-\la)^{-1}$. On the other side, asymptotic expansions of the second term, $g(x,\la,\mu)$, in (\ref{g8}) is independent on the order, since substituting this equality in (\ref{g28}) we get:
\begin{align}
g(x,\la,\mu)&=-\dfrac{1}{2\pi{i}}\int\dfrac{d\bar{z}{\wedge}dz\,r(z)}{(\la-\bar{z})(\mu-z)}e^{(\ell(\bar{z})-\ell(z))x}-\nonumber\\
&-\dfrac{1}{(2\pi{i})^2}\int\dfrac{d\bar{z}{\wedge}dz\,r(\bar{z})}{\la-z}\int\dfrac{d\bar{z'}{\wedge}dz'\,r(z')}{\mu-z'}e^{(\ell(z)-\ell(\bar{z})+\ell(\bar{z'})-\ell(z'))x}
f(x,\bar{z},\bar{z'}).\label{h32}
\end{align}
We introduce notation $g(x,l,m)$, $l,m=0,1,2,\ldots$, for the coefficients of asymptotic expansion of this function when $\la$ and/or $\mu$ tend to infinity:
\begin{equation}
g(x,\la,\mu)=\sum_{l=0}^{\infty}\dfrac{g(x,l,\mu)}{\la^{l+1}}=\sum_{m=0}^{\infty}\dfrac{g(x,\la,m)}{\mu^{m+1}}=\sum_{l,m=0}^{\infty}\dfrac{g(x,l,m)}{\la^{l+1}\mu^{m+1}},\label{as1}
\end{equation}
so that
\begin{equation}
g(x,l,\mu)=\sum_{m=0}^{\infty}\dfrac{g(x,l,m)}{\mu^{m+1}},\qquad g(x,\la,m)=\sum_{l=0}^{\infty}\dfrac{g(x,l,m)}{\la^{l+1}}.\label{as4}
\end{equation}
Thanks to (\ref{g5})--(\ref{g7}) we have that
\begin{align}
g(x,0,\mu)&=1-\chi(x,\mu),\label{as5}\\
g(x,\la,0)&=-1+\xi(x,\la).\label{as6}\\
g(x,0,0)=w(x),\label{as7}
\end{align}
where it is necessary to underline that, say $g(x,0,\mu)$ is not the value of $g(x,\la,\mu)$ at $\la=0$, but the coefficient of $1/\la$ in the first equality in (\ref{as1}).

Because of the above the last term in the r.h.s.\ of  (\ref{h29}) is independent on the order of the asymptotic expansions. The remarkable fact is that the first two terms in the r.h.s.\ of this equality are also independent of the order of operations, since by (\ref{as1}) and (\ref{as4}) we have that:
\begin{align}
&\dfrac{g(x,\la,\mu)-g(x,\la,\nu)}{\mu-\nu}=-\sum_{l=0}^{\infty}\sum_{m=0}^{\infty}\sum_{k=0}^{\infty}\dfrac{g(x,l,k+m)}{\la^{l+1}\mu^{m+1}\nu^{k+1}},\label{as13}\\
&\dfrac{g(x,\la,\mu)-g(x,\nu,\mu)}{\la-\nu}=-\sum_{l=0}^{\infty}\sum_{m=0}^{\infty}\sum_{k=0}^{\infty}\dfrac{g(x,l+k,m)}{\la^{l+1}\mu^{m+1}\nu^{k+1}}.\label{as14}
\end{align}
Thus asymptotic series of the r.h.s.\ of (\ref{as9}), that is just another form of  (\ref{h29}),  are independent of the order three complex parameters $\la$, $\mu$ and $\nu$ tend to infinity. Eq.\ (\ref{as9}) is more convenient for getting the asymptotic expansion, while intermediate objects need special consideration. Following notation (\ref{as1}) we introduce
\begin{equation}
f(x,\la,\mu)=\left\{\begin{array}{lr}
\displaystyle\sum_{l=0}^{\infty}\dfrac{f'(x,l,\mu)}{\la^{l+1}},&\la\to\infty,\\
\displaystyle\sum_{m=0}^{\infty}\dfrac{f''(x,\la,m)}{\mu^{m+1}},&\mu\to\infty, \end{array}\right.\label{h6}
\end{equation}
where because of (\ref{as4})
\begin{equation}
f'(x,l,\mu)=-\mu^{l}+g(x,l,\mu),\qquad f''(x,\la,m)=\la^{m}+g(x,\la,m),\quad l,m=0,1,2,\ldots.\label{h7}
\end{equation}
So we get different and growing expressions for these coefficient functions.

The Jost solutions were defined in (\ref{g5}) and (\ref{g6}) (see also (\ref{as5}) and (\ref{as6})) as limiting values of the Cauchy--Jost function, that in these terms sounds as
\begin{equation}
f'(x,0,\mu)=-\chi(x,\mu),\qquad f''(x,\la,0)=\xi(x,\la).\label{h8}
\end{equation}
Since the asymptotic expansion of (\ref{as9}) does not depend on the order we can use this equality in order to get time evolutions of the Jost solutions and potential, that justifies (\ref{h24}), (\ref{h25}) and (\ref{h28}) above. Thus by (\ref{h6}) we derive from (\ref{as9}):
\begin{align}
&X(\nu)f'(x,l,\mu)=\dfrac{f'(x,l,\mu)}{\mu-\nu}-\sum_{j=0}^{l-1}\nu^{l-1-j}f'(x,j,\mu)-f'(x,l,\nu)f(x,\nu,\mu),\label{h18}\\
&X(\nu)f''(x,\la,m)=\sum_{j=0}^{m-1}\nu^{m-1-j}f''(x,\la,j)-\dfrac{f''(x,\la,m)}{\la-\nu}-f(x,\la,\nu)f''(x,\nu,m).\label{h19}
\end{align}
Using (\ref{h7}) we can get asymptotic decompositions of these equalities at $\nu\to\infty$:
\begin{align}
&\partial_{x_k}f'(x,l,\mu)=f'(x,l+k,\mu)-\mu^{k}f'(x,l,\mu)-\sum_{j=0}^{k-1}g(x,l,k-1-j)f'(x,j,\mu),\label{h20}\\
&\partial_{x_k}f''(x,\la,m)=-f''(x,\la,m+k)+\la^{k}f''(x,\la,m)-\sum_{j=0}^{m-1}f''(x,\la,k-1-j)g(x,j,m).\label{h21}
\end{align}
On the other side, decomposing (\ref{as9}) with respect to $\nu$ first, we get
\begin{equation}
\partial_{x_k}f(x,\la,\mu)=(\la^{k}-\mu^{k})f(x,\la,\mu)-\sum_{l=0}^{k-1}f''(x,\la,k-l-1)f'(x,l,\mu),\label{h9}
\end{equation}
and it is easy to show that because of (\ref{h7}) all positive powers of $\la$ and $\mu$ cancel out, and the asymptotic decompositions of these equalities with respect to these variables give (\ref{h20}) and (\ref{h21}) correspondingly. We see that in the case where in (\ref{h18}) $l=0$ (or in (\ref{h19}) $m=0$) the r.h.s.'s are simplified and thanks to (\ref{h8}) we arrive to (\ref{h24}) and (\ref{h25}). Further asymptotic expansions of (\ref{h20}) and (\ref{h21}) is possible after substitution of (\ref{h7}) that supplies
\begin{equation}
\partial_{x_k}g(x,l,m)=g(x,l+k,m)-g(x,l,k+m)-\sum_{j=0}^{k-1}g(x,l,j)g(x,k-1-j,m).\label{as15}
\end{equation}
Setting here $l=m=0$ we get (\ref{h28}) because of notation (\ref{as5})--(\ref{as7}).  Let us mention that the term of order $\nu^{-1}$ of the expansion of (\ref{h28}) gives zero, terms $\nu^{-2}$ and $\nu^{-3}$  give identities, while for the term  $\nu^{-4}$ we get
\begin{equation}
8w_{x_3}(x)=2w_{x_1x_1x_1}(x)+12(w_{x_1}(x)^2)+6\partial_{x_1}^{-1}w_{x_2x_2}(x),\label{as161}
\end{equation}
i.e., the mKPII equation that thanks to (\ref{old4}) takes the standard form of the KPII:
\begin{equation}
(u_{t}-3(u^{2})_{x_1}+u_{x_1x_1x_1})_{x_1}=-3u_{x_2x_2}.\label{as162}
\end{equation}

Eq.\ (\ref{as15}) gives also the recursion procedure to determine all coefficients $g(x,l,m)$ starting with $g(x,0,0)$ in (\ref{as7}). Indeed, the highest coefficients in the r.h.s. of this equality enter as difference $g(x,l+k,m)-g(x,l,k+m)$. On the other side for $k=1$ we have
\begin{equation}
\partial_{x_1}g(l,m)=g(l+1,m)-g(l,m+1)-g(l,0)g(0,m),\label{as17}
\end{equation}
so that the same coefficients will appear in $\partial_{x_1}\partial_{x_{k-1}}g(x,l,m)$ as a sum $g(x,l+k,m)+g(x,l,k+m)$.

Let us consider special cases of (\ref{h9}) corresponding to the lowest evolutions $x_1$, $x_2$ and $x_3$:
\begin{align}
\partial_{x_1}f(x,\la,\mu)&=(\la-\mu)f(x,\la,\mu)-f''(x,\la,0)f'(x,0,\mu),\label{h10}\\
\partial_{x_2}f(x,\la,\mu)&=(\la^2-\mu^2)f(x,\la,\mu)-f''(x,\la,1)f'(x,0,\mu)-f''(x,\la,0)f'(x,1,\mu),\label{h11}\\
\partial_{x_3}f(x,\la,\mu)&=(\la^{3}-\mu^{3})f(x,\la,\mu)-\nonumber\\
&-f''(x,\la,2)f'(x,0,\mu)-f''(x,\la,1)f'(x,1,\mu)-f''(x,\la,0)f'(x,2,\mu),\label{h111}
\end{align}
In terms of notation (\ref{as7}) we derive from (\ref{h20}) and (\ref{h21})
\begin{align}
&\partial_{x_1}f'(x,0,\mu)=-(\mu+w(x))f'(x,0,\mu)+f'(x,1,\mu),\label{h12}\\
&\partial_{x_1}f''(x,\la,0)=(\la-w(x))f''(x,\la,0)-f''(x,\la,1),\label{h13}
\end{align}
and then
\begin{align}
&\Lo_{0}(x,\partial_{x})f(x,\la,\mu)=-2\mu f_{x_1}(x,\la,\mu)-2f''_{x_1}(x,\la,0)f'(x,0,\mu),\label{h14}\\
&\Lo^{\text{d}}_{0}(x,\partial_{x})f(x,\la,\mu)=2\la f_{x_1}(x,\la,\mu)-2f''(x,\la,0)f'_{x_1}(x,0,\mu).\label{h15}
\end{align}
These equalities give action of differential operators on $f(x,\la,\mu)$ in terms of objects with one 0, i.e., in terms of the Jost solutions by (\ref{h8}):
\begin{align}
&\Lo_{0}(x,\partial_{x})f(x,\la,\mu)=-2\mu f_{x_1}(x,\la,\mu)+2\xi_{x_1}(x,\la)\chi(x,\mu),\label{h16}\\
&\Lo^{\text{d}}_{0}(x,\partial_{x})f(x,\la,\mu)=2\la f_{x_1}(x,\la,\mu)+2\xi(x,\la)\chi_{x_1}(x,\mu),\label{h17}
\end{align}
so that the leading terms of the expansion of the first equality with respect to $\la$ and the second one with respect to $\mu$ are exactly (\ref{old431}) by (\ref{g5})--(\ref{g7}).

\section{Conclusion}
Summarizing, we see that equation (\ref{as9}) together with the asymptotic behavior given in (\ref{g234}) and representation (\ref{g8}) defines all constituents of the KPII hierarchy together with their evolution. In their turn all these evolutions of the potential are given in a compact form by  (\ref{h28}).

\noindent\textbf{Acknowledgment.} Sections 3, 4 and 5 of the article are performed by A.\ K.\ Pogrebkov, sections 1, 2 and 6 are performed by M.\ Boiti and F.\ Pempinelli. Investigation of A.\ K.\ Pogrebkov is carried out at the expense of the Russian Science Foundation (grant $\No$ 14-50-00005) in Steklov Mathematical Institute of Russian Academy of Sciences.

\end{document}